\documentclass[10pt,journal,compsoc]{IEEEtran}

\usepackage[utf8]{inputenc} 
\usepackage[T1]{fontenc}    
\usepackage{hyperref}       
\usepackage{url}            
\usepackage{booktabs}       
\usepackage{amsfonts}       
\usepackage{nicefrac}       
\usepackage{microtype}      
\usepackage{xcolor}         

\usepackage{subcaption}
\usepackage[font={small}]{caption}

\usepackage[pdftex]{graphicx} 
\usepackage{tikz} 
\usepackage{pgfplots} 
\pgfplotsset{compat=newest}
\usetikzlibrary{spy, calc}

\definecolor{color0}{HTML}{75b9be}
\definecolor{color1}{HTML}{ee7674}
\definecolor{color2}{HTML}{832161}
\definecolor{color3}{HTML}{eab464}
\definecolor{color4}{HTML}{335145}

\usepackage{amsmath}
\usepackage{mathtools} 
\usepackage{amssymb} 
\usepackage{amsthm} 
\usepackage{thmtools} 
\usepackage{bm} 
\usepackage{csquotes} 
\usepackage{balance} 
\ifCLASSOPTIONcompsoc
  \usepackage[nocompress]{cite}
\else
  \usepackage{cite}
\fi
\usepackage{ctable} 
\usepackage{enumitem} 
\usepackage{wrapfig}

\usepackage{algorithm}
\usepackage{algpseudocode}

\usepackage{siunitx}
\makeatletter
\@ifundefined{numproduct}
{%
  \newcommand{\numproduct}[1]{\num{#1}}
}
{%
}%
\makeatother

\usepackage[capitalize]{cleveref}
\crefname{equation}{\unskip}{\unskip}
\crefname{appsec}{Appendix}{Appendices}

\newlength\figureheight
\newlength\figurewidth
\newlength\figureheightc
\newlength\figurewidthc
\pgfplotsset{
  compat=1.11,
  legend image code/.code={
  \draw[mark repeat=2,mark phase=2]
    plot coordinates {
    (0cm,0cm)
    (0.15cm,0cm)        
    (0.3cm,0cm)         
  };%
  }
}

\newtheorem{example}{Example}

\newcommand{\transpose}[1]{#1^{\textup{\textsf{T}}}}

\newcommand{\define}{\triangleq}
\DeclareMathOperator*{\argmax}{arg\,max}
\DeclareMathOperator*{\argmin}{arg\,min}

\newcommand{\F}{\mathbb F} 
\newcommand{\R}{\mathbb R} 
\newcommand{\N}{\mathbb N} 
\newcommand{\sL}{\mathcal L} 

\newcommand{\Nn}{N} 

\newcommand{\bX}{\bm X} 
\newcommand{\bx}{\bm x} 
\newcommand{\bY}{\bm Y} 
\newcommand{\bH}{\bm H} 

\newcommand{\bV}{\bm V}

\begin{document}

\title{DSAG: A mixed synchronous-asynchronous iterative method for straggler-resilient learning}

\author{Albin~Severinson,~\IEEEmembership{Student Member,~IEEE,}
        Eirik~Rosnes,~\IEEEmembership{Senior Member,~IEEE,}
        Salim~El~Rouayheb,~\IEEEmembership{Senior Member,~IEEE,} and
        Alexandre~Graell~i~Amat,~\IEEEmembership{Senior Member,~IEEE}
\IEEEcompsocitemizethanks{
\IEEEcompsocthanksitem The work of A. Graell i Amat was supported by the Swedish Research Council under grant 2020-03687.\protect\\
\IEEEcompsocthanksitem A. Severinson is with Simula UiB and the Department of Informatics at the University of Bergen, N-5020 Bergen, Norway (email: albin@simula.no).\protect\\
\IEEEcompsocthanksitem E. Rosnes is with Simula UiB, N-5006 Bergen, Norway (email: eirikrosnes@simula.no).\protect\\
\IEEEcompsocthanksitem S. El Rouayheb is with the Department of Electrical and Computer Engineering, Rutgers University, 08854 Piscataway, New Jersey (email:  salim.elrouayheb@rutgers.edu).\protect\\
\IEEEcompsocthanksitem A. Graell i Amat is with the Department of Electrical Engineering, Chalmers University of Technology, SE-41296 Gothenburg, Sweden (email: alexandre.graell@chalmers.se) and Simula UiB, N-5006 Bergen, Norway.}
}%

%

\IEEEtitleabstractindextext{%
\begin{abstract}
  We consider straggler-resilient learning. In many previous works, e.g., in the coded computing literature, straggling is modeled as random delays that are independent and identically distributed between workers. However, in many practical scenarios, a given worker may straggle over an extended period of time. We propose a latency model that captures this behavior and is substantiated by traces collected on Microsoft Azure, Amazon Web Services (AWS), and a small local cluster. Building on this model, we propose DSAG, a mixed synchronous-asynchronous iterative optimization method, based on the stochastic average gradient (SAG) method, that combines timely and stale results. We also propose a dynamic load-balancing strategy to further reduce the impact of straggling workers. We evaluate DSAG for principal component analysis, cast as a finite-sum optimization problem, of a large genomics dataset, and for logistic regression on a cluster composed of \num{100} workers on AWS, and find that DSAG is up to about \num{50}\% faster than SAG, and more than twice as fast as coded computing methods, for the particular scenario that we consider.
\end{abstract}
}

\maketitle

\IEEEdisplaynontitleabstractindextext

%
\IEEEpeerreviewmaketitle

\IEEEraisesectionheading{\section{Introduction}\label{sec:introduction}}

%
%
%
%

\IEEEPARstart{W}{e} are interested in reducing the latency of distributed iterative optimization methods for empirical risk minimization. In particular, we want to reduce the impact of straggling workers, i.e., workers experiencing delays, which can significantly slow down distributed algorithms. The straggler problem is a consequence of the design of modern large-scale compute clusters (sometimes referred to as \emph{warehouse-scale computers}), which are built from a large number of commodity servers connected in a heterogeneous manner, and where many virtual machines may share the same physical host server, to maximize cost-efficiency \cite{Dean2013,Barroso2018}. Examples include Microsoft Azure, Google Cloud, and Amazon Web Services (AWS).

Straggling is often assumed, e.g., in the coded computing literature \cite{Lee2017,Tandon2017,Karakus2017,Yang2018}, to be caused by random delays that are independent and identically distributed (i.i.d.) between workers and iterations. However, from traces collected on Microsoft Azure and AWS, we find that stragglers tend to remain stragglers. As a result, data processed by stragglers may never factor in for stochastic methods that only rely on the results from the fastest subset of workers.

This work consists of three parts. First, we propose a latency model that, unlike previous models, accounts for differences in the mean and variance of the latency between different workers and over time. Further, for the proposed model we show how to efficiently estimate the latency of the $w$-th fastest worker out of a set of $\Nn$ workers, including for iterative computations, where a worker may remain unavailable over several subsequent iterations.

Second, based on this model, we propose DSAG, an iterative method for finite-sum optimization (machine learning problems are typically cast as finite-sum optimization problems) which adapts the stochastic average gradient (SAG) method \cite{Schmidt2017} to distributed environments. The key idea of DSAG is to wait for the $w$ fastest workers in each iteration\textemdash i.e., DSAG is a stochastic method\textemdash while simultaneously integrating stale results received from the $\Nn-w$ stragglers as they are received over subsequent iterations. DSAG relies on the \emph{variance reduction} technique of SAG to suppress the potentially high variance caused by this strategy and improve convergence. Finally, we propose a dynamic load-balancing strategy for reducing the variation in latency between workers, that is based on the model proposed in part one.

We validate the proposed model on Azure,  AWS, and a small local cluster, and find that the model accurately predicts latency across the three platforms. We evaluate the performance of DSAG by using it for principal component analysis (PCA), cast as an optimization problem, of a large genomics dataset, and for logistic regression. For both PCA and
logistic regression, DSAG with load balancing reduces
latency significantly compared to SAG\textemdash for a scenario with 100 workers on AWS, DSAG is about 10\% faster than SAG for PCA and up to 50\% faster  for logistic regression. Furthermore, it is more than twice as fast as coded computing methods.

We provide the source code of our implementation and the latency traces we have collected under a permissive license at~\cite{dsagsource}.

\subsection*{Related work}

Recently, \emph{coded computing} has been proposed to deal with stragglers \cite{Lee2017}. The key idea is to add redundant computations (thus increasing the per-worker computational load), such that the result of the computation can be recovered from a subset of the workers, typically via a decoding operation. Coded computing methods have been proposed for, e.g., matrix-vector multiplication~\cite{Lee2017,Severinson2019tcom,Severinson2018}, matrix-matrix multiplication~\cite{Yu2017,Yu2018isit,Dutta2018,Lee2017isit,Fahim2019isit,Dutta2020,Gupta2018,Jahani-Nezhad2019}, polynomial evaluation~\cite{Yang2019}, and gradient computations~\cite{Tandon2017,Ye2018}. For example, the method in~\cite{Tandon2017} increases the computational load per worker by a factor $(\Nn-w)+1$ compared to gradient descent (GD) to tolerate any $\Nn-w$ stragglers.

Another method to deal with stragglers is stochastic optimization, the simplest form of which is to ignore stragglers for GD. This is a stochastic gradient descent (SGD) method, sometimes referred to as ignoring stragglers SGD. SGD does not converge to the optimum unless the stepsize is reduced as the algorithm progresses. However, a smaller stepsize reduces the rate of convergence, and it is difficult to determine the correct rate at which to reduce the stepsize. Approximate coded computing methods combine ignoring stragglers SGD with redundancy, e.g.,~\cite{Karakus2017,Bitar2020,Wang2019}. These methods improve the rate of convergence per iteration compared to ignoring stragglers SGD but typically do not converge to the optimum, and typically increase the computational load compared to GD by a factor $2$ or $3$.

The above methods treat iterations independently, ignoring the correlation between the results computed in subsequent iterations, which is often significant. The coded version of the power method proposed in~\cite{Yang2018} is an exception in that the previous iterate is used as side information during decoding. The process is related to \emph{sketch-and-project} methods (see, e.g., \cite{Richtarik2020,Hanzely2018,Gower2015}), i.e., iterative methods to approximate some quantity from low-rank sketches. In particular, the method in~\cite{Yang2018} can be seen as a special case of the one in~\cite{Hanzely2018}. A significant shortcoming of the method in~\cite{Yang2018} is that it requires a complex decoding process to be performed by the coordinator for each iteration.

The method in~\cite{Hanzely2018} is a variance-reduced stochastic method for first-order optimization. For each iteration, these methods use an estimate of the gradient to, e.g., perform a gradient step, i.e., they are stochastic. Variance-reduced methods converge to the optimum despite being stochastic by using information contained in previous iterates and/or gradients to ensure that the variance of this estimate tends to zero as the method progresses. Examples of variance-reduced methods include SAG~\cite{Schmidt2017}, SAGA~\cite{Defazio2014} (including a peer-to-peer version~\cite{Calauzenes2017}), SARAH~\cite{Nguyen2017}, SVRG~\cite{Johnson2013}, SEGA~\cite{Hanzely2018}, and MARINA~\cite{Gorbunov2021}. These works do not consider the straggler problem.

Exploiting stale gradients in combination with asynchronicity to alleviate the straggler problem has been explored in several previous works, see, e.g., \cite{Lia15,Gup16}, and references therein, in the neighboring area of federated learning, and~\cite{Dutta2018}.  These methods are similar to the proposed DSAG, but do not employ variance reduction. For example, a mixed synchronous-asynchronous distributed version of SGD that is similar to ours has been proposed and analyzed in~\cite{Dutta2018}. Like the method we propose, the method in~\cite{Dutta2018} uses asynchronicity to reduce iteration latency. However, unlike our method, the method in~\cite{Dutta2018} gradually increases the level of synchronicity, thus increasing iteration latency, to improve convergence, whereas our method relies on variance reduction. There has also been a significant amount of work on asynchronous optimization for shared-memory systems, e.g., \cite{Recht2011,Xinghao2016}, and references therein. However, these works do not consider the straggler problem.

The load-balancing approach we suggest is designed specifically for DSAG, but is inspired by the large number of previous works on the topic; see, e.g.,~\cite{Catalyurek2007,Merrill2016,Javadi2017,Maji2015}, and references therein. These suggest approaches to balance either i) the complexity of the subtasks that make up a particular large computation (e.g.,~\cite{Catalyurek2007,Merrill2016}), or ii) incoming requests between instances of a distributed application, such as a web server (e.g.,~\cite{Javadi2017,Maji2015}). The approach we suggest, like those of~\cite{Javadi2017,Maji2015}, but unlike~\cite{Catalyurek2007,Merrill2016}, accounts for latency differences between servers and over time\textemdash as is the case in the cloud\textemdash but balances the complexity of subtasks, as in~\cite{Catalyurek2007,Merrill2016}. Furthermore, DSAG is designed with load-balancing in mind, and, as a result, unlike the approach of~\cite{Catalyurek2007,Merrill2016}, does not require moving data between servers to perform load-balancing.

\section{Preliminaries}
\label{sec:preliminaries}

Denote by $\bX \in \R^{n \times d}$ a data matrix, where $n$ is the number of samples and $d$  the dimension. Many learning problems (e.g., linear and logistic regression, PCA, matrix factorization, and training neural networks) can be cast as a finite-sum optimization problem of the form
\begin{equation} \label{eq:optimization}
  \bV^* = \argmin_{\bV \in \sL} \left[ F(\bV, \bX) \define R(\bV) + \sum_{i=1}^n f_i(\bV, \bx_i) \right],
\end{equation}
where $\sL$ is the solution space, $f_i$ is the loss function with respect to the $i$-th sample (row of $\bX$), which we denote by $\bx_i$, and $R$ is a regularizer, which serves to, e.g., bias $\bV^*$ toward sparse solutions. For the remainder of this paper, we write $F(\bV)$ and $f_i(\bV)$, leaving the dependence of $F$ and $f_i$ on $\bX$ and $\bx_i$, respectively, implicit.

These problems are often solved (e.g., for the examples mentioned above) using so-called first-order iterative optimization methods, i.e., methods that iteratively update a solution based on the gradient of the loss function $F$, which we denote by $\nabla F$. One example of such a method is GD, the update rule of which is
\begin{equation} \label{eq:pcagd}
  \bV^{(t+1)} = G\left( \bV^{(t)} - \eta \nabla F\left(\bV^{(t)}\right) \right),
\end{equation}
where $t$ is the iteration index, $\eta$ the stepsize, and $G$ a projection operator (possibly the identity operator).

In this work, we consider a distributed scenario in which the rows of $\bX$ are stored over $\Nn$ worker nodes, 
such that each node stores an equal fraction of the rows. The workers are responsible for computing the subgradients $\nabla f_1, \dots, \nabla f_n$ and the coordinator is responsible for aggregating those subgradients and performing a gradient step.

\subsection*{Experimental setup}
\label{sec:computing}

The results presented in this work are from experiments conducted on compute clusters hosted on Microsoft Azure (region \texttt{West Europe}), AWS (region \texttt{eu-north-1}), and the eX3 cluster.\footnote{See~\url{ex3.simula.no}.} For Azure, the nodes are of type \texttt{F2s\_v2} and for AWS the nodes are of type \texttt{c5.xlarge}.\footnote{Both \texttt{F2s\_v2} and \texttt{c5.xlarge} nodes are based on Intel Xeon Platinum 8000 series processors. \texttt{F2s\_v2} nodes have an expected network speed of \num{875} Mbps, whereas \texttt{c5.xlarge} nodes have a network speed of up to $10$ Gpbs.} For AWS, we also provide traces for nodes of type \texttt{c5.xlarge} in region \texttt{us-east-1} and of type \texttt{t3.xlarge} in region \texttt{eu-north-1}~\cite{dsagsource}. The nodes used on eX3 are equipped with AMD EPYC 7302P processors and high-speed InfiniBand interconnects. We use the same type of node for the coordinator and the workers. Our implementation is written in the Julia programming language,
and we use OpenMPI for communication\textemdash specifically, the \texttt{Isend} and \texttt{Irecv} nonblocking, point-to-point communication subroutines. For Azure and AWS, we use the CycleCloud and ParallelCluster systems, respectively, to create workers on-demand.

Throughout this paper, we consider a data matrix derived from the $1000$ Genomes phase-$3$ dataset~\cite{Auton2015}. More precisely, we consider a binary representation of the data for each chromosome, where a nonzero entry in the $(i, j)$-th position indicates that the genome of the $i$-th subject differs from that of the reference genome in the $j$-th position. The matrix we use is the concatenation of such matrices computed for each chromosome. It is a sparse matrix of size \numproduct{81271767x2504} with density about \num{5.360388070027386}\%. 
In~\cref{sec:convergence}, we also consider the HIGGS dataset, which consists of \num{11000000} samples with \num{28} features~\cite{Baldi2014}. For all computations, each worker stores the subset of the dataset assigned to it in memory throughout the computation.

\section{Modeling the latency of gradient computations}
\label{sec:latencymodel}

\begin{figure}[t]
  \vskip -5mm
  \begin{tikzpicture}

\definecolor{color0}{HTML}{75b9be}
\definecolor{color1}{HTML}{ee7674}
\definecolor{color2}{HTML}{832161}
\definecolor{color3}{HTML}{eab464}
\definecolor{color4}{HTML}{335145}
    
\begin{axis}[
name=plot1,
legend cell align={left},
legend style={font=\scriptsize, fill opacity=0.8, draw opacity=1, text opacity=1, at={(0.98,0.02)}, anchor=south east, draw=white!80!black},
tick pos=left,
grid style={gray,opacity=0.5,dotted},
xmajorgrids,
xticklabels=\empty,
height=\figureheight,
width=\figurewidth,
xmin=1e6, xmax=1e9,
ylabel={Computation latency mean [s]},
ymajorgrids,
xmode=log,
ymode=log,
ymin=1e-3, ymax=1e0,
]

\addplot [only marks, color1, mark=*, mark size=1, mark options={fill=white,solid}, forget plot]
table {./figures/linear_latency/data/aws/comp_mean.csv};

\addplot [only marks, color0, mark=*, mark size=1, mark options={fill=white,solid}, forget plot]
table {./figures/linear_latency/data/azure/comp_mean.csv};

\addplot [only marks, color1, mark=square*, mark size=2, mark options={fill=white,solid}]
table {./figures/linear_latency/data/aws/comp_mean_mean.csv};
\addlegendentry{AWS}

\addplot [semithick, color1, forget plot]
table {./figures/linear_latency/data/aws/comp_mean_line.csv};

\addplot [only marks, color0, mark=square*, mark size=2, mark options={fill=white,solid}]
table {./figures/linear_latency/data/azure/comp_mean_mean.csv};
\addlegendentry{Azure}

\addplot [semithick, color0, forget plot]
table {./figures/linear_latency/data/azure/comp_mean_line.csv};

\end{axis}     
\begin{axis}[
name=plot2,
at=(plot1.below south west), anchor=above north west,
legend cell align={left},
legend style={font=\scriptsize, fill opacity=0.8, draw opacity=1, text opacity=1, at={(0.98,0.02)}, anchor=south east, draw=white!80!black},
tick pos=left,
grid style={gray,opacity=0.5,dotted},
xlabel style={align=center}, xlabel={Computational load ($c$)},
xmajorgrids,
height=\figureheight,
width=\figurewidth,
xmin=1e6, xmax=1e9,
ylabel={Computation latency variance [$s^2$]},
ymajorgrids,
xmode=log,
ymode=log,
ymin=1e-8, ymax=1e-4,
]

\addplot [only marks, color1, mark=*, mark size=1, mark options={fill=white,solid}]
table {./figures/linear_latency/data/aws/comp_var.csv};

\addplot [only marks, color0, mark=*, mark size=1, mark options={fill=white,solid}]
table {./figures/linear_latency/data/azure/comp_var.csv};

\addplot [only marks, color1, mark=square*, mark size=2, mark options={fill=white,solid}]
table {./figures/linear_latency/data/aws/comp_var_mean.csv};

\addplot [semithick, color1]
table {./figures/linear_latency/data/aws/comp_var_line.csv};

\addplot [only marks, color0, mark=square*, mark size=2, mark options={fill=white,solid}]
table {./figures/linear_latency/data/azure/comp_var_mean.csv};

\addplot [semithick, color0]
table {./figures/linear_latency/data/azure/comp_var_line.csv};

\end{axis}       
\end{tikzpicture}
  \caption{Mean and variance of the computation latency recorded for $100$ different workers as a function of the computational load. Circles correspond to workers, and the mean over all recordings for each computational load is marked by a square. For reference, we also show a line passing through the origin fitted to the data.}
  \label{fig:linear_latency}
  \vskip -5mm
\end{figure}
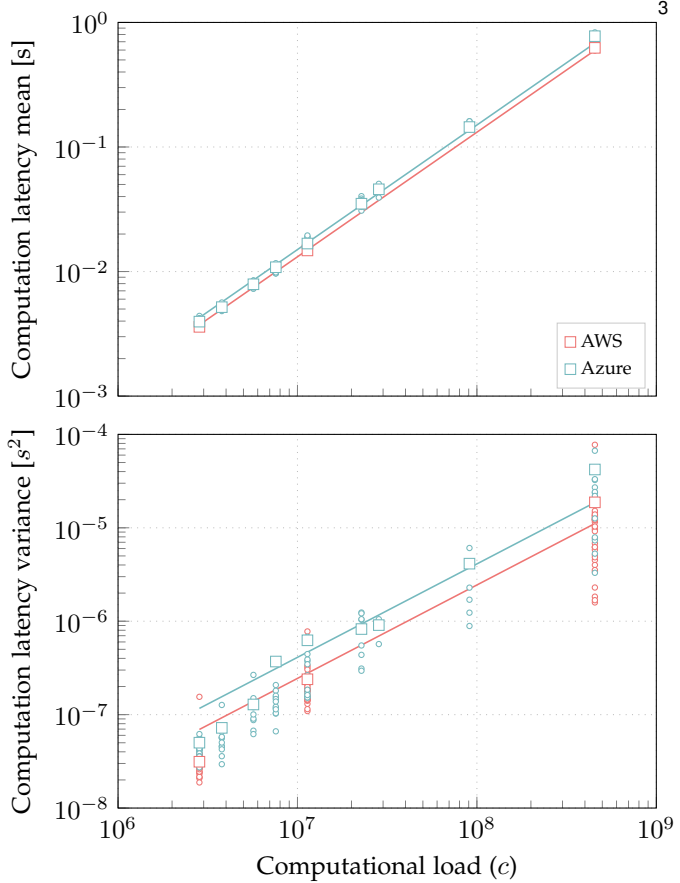

In this section, we propose a model of the communication and computation latency of workers performing gradient computations in a distributed setting. Later, we use this model to predict the latency of the $w$-th fastest worker out of a set of workers. We first consider the latency of workers operating in \emph{steady state} (\cref{sec:steadystate}), after which we consider how the latency of a particular worker changes over time (\cref{sec:statetransition}).

The model we propose is based on latency traces collected in clusters composed of up to \num{108} workers on AWS, Azure, and eX3, with varying per-worker computational load, which we denote by $c$, and $b$ bytes communicated per iteration. Here, the computational load can be any quantity that captures the amount of work performed by each worker and iteration, such that a change in $c$ results in a proportional change in the expected computation latency of a single worker. The number of bytes communicated and the computational load are equal for all workers, and we repeat the experiment at different days and times of the day.

In particular, for each worker, we record the latency associated with sending to the worker an iterate $\bV$ and for the worker to respond with the result of the computation,
\begin{equation} \label{eq:kernel}
  \transpose{\bX_{i:j}} \bX_{i:j} \bV,
\end{equation}
for some integers $1 \leq i \leq j \leq n$, 
where $\bX_{i:j}$ denotes the submatrix of $\bX$ consisting of rows $i$ through $j$. Hence, our results generalize to computations that rely on matrix multiplication, although the model is also easily adapted to other types of computations. In addition, we make available traces recorded for other computations and datasets, and we find consistent behavior across the computations and datasets considered~\cite{dsagsource}.

We let $c$ be the number of operations required to perform this computation, i.e., $c=2\zeta dk (j - i + 1)$, where $d$ is the dimension, $k$ is the number of columns of $\bV$, and $\zeta$ is the density of the data matrix. For all recordings, we randomly permute the rows of the matrix to break up dense blocks, and we adjust the computational load by tuning the number of samples processed.

In~\cref{fig:linear_latency}, we plot the range of computational loads considered, together with the mean and variance of the computation latency recorded for $100$ different workers for each computational load, when the number of bytes communicated per iteration is $b=\num{30048}$. For reference, we also plot a line passing through the origin fitted to the data.

\subsection{Steady-state latency}
\label{sec:steadystate}

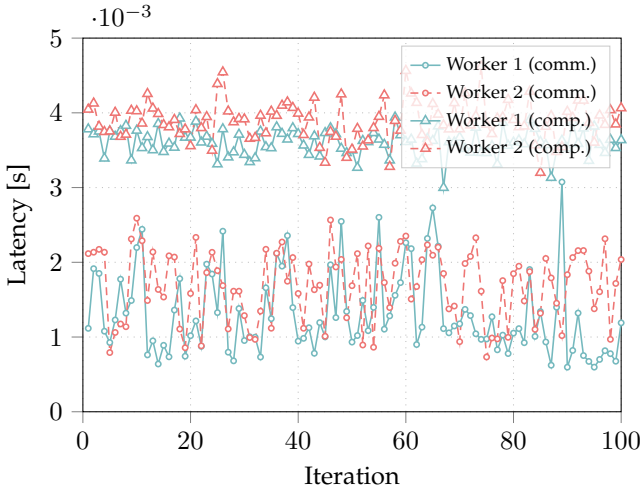
\begin{figure}[t]
\begin{tikzpicture}

\definecolor{color0}{HTML}{75b9be}
\definecolor{color1}{HTML}{ee7674}
\definecolor{color2}{HTML}{832161}
\definecolor{color3}{HTML}{eab464}
\definecolor{color4}{HTML}{335145}

\pgfplotsset{
legend image code/.code={
\draw[mark repeat=2,mark phase=2]
plot coordinates {
(0cm,0cm)
(0.2cm,0cm)        
(0.4cm,0cm)         
};%
}
}

\begin{axis}[
name=plot1,
legend cell align={left},
legend style={font=\scriptsize, fill opacity=0.8, draw opacity=1, text opacity=1, at={(0.98,0.98)}, anchor=north east, draw=white!80!black},
tick align=outside,
tick pos=left,
grid style={gray,opacity=0.5,dotted},
xlabel={Iteration},
xmajorgrids,
height=\figureheight,
width=\figurewidth,
xmin=0, xmax=100,
ylabel={Latency [s]},
ymajorgrids,
ymin=0, ymax=5e-3,
ytick distance=1e-3,
]

\addplot [semithick, color0, mark=*, mark size=1, mark options={fill=white}] table {./figures/timeseries/data/Azure/timeseries_communication_39_1.csv};
\addlegendentry{Worker $1$ (comm.)}
\addplot [semithick, color1, densely dashed, mark=*, mark size=1, mark options={fill=white, solid}] table {./figures/timeseries/data/Azure/timeseries_communication_39_11.csv};
\addlegendentry{Worker $2$ (comm.)}

\addplot [semithick, color0, mark=triangle*, mark size=2, mark options={fill=white}] table {./figures/timeseries/data/Azure/timeseries_compute_39_1.csv};
\addlegendentry{Worker $1$ (comp.)}
\addplot [semithick, color1, densely dashed, mark=triangle*, mark size=2, mark options={fill=white, solid}] table {./figures/timeseries/data/Azure/timeseries_compute_39_11.csv};
\addlegendentry{Worker $2$ (comp.)}

\end{axis}

\end{tikzpicture}
  \caption{Per-iteration latency of two workers on Azure, with $b=\num{30048}$ bytes communicated per iteration (circles) and computational load $c=\num{2.840789371049846e6}$ (triangles).}  
  \label{fig:timeseries}  
  \vspace{-5mm}
\end{figure}%
\begin{figure}[t]
  \raisebox{3.25mm}{
\begin{tikzpicture}

\definecolor{color0}{HTML}{75b9be}
\definecolor{color1}{HTML}{ee7674}
\definecolor{color2}{HTML}{832161}
\definecolor{color3}{HTML}{eab464}
\definecolor{color4}{HTML}{335145}

\pgfplotsset{
legend image code/.code={
\draw[mark repeat=2,mark phase=2]
plot coordinates {
(0cm,0cm)
(0.2cm,0cm)        
(0.4cm,0cm)         
};%
}
}

\begin{axis}[
name=plot1,
legend cell align={left},
legend style={font=\scriptsize, fill opacity=0.8, draw opacity=1, text opacity=1, at={(0.98,0.02)}, anchor=south east, draw=white!80!black},
tick align=outside,
tick pos=left,
grid style={gray,opacity=0.5,dotted},
ylabel={CDF},
xmajorgrids,
height=\figureheight,
width=0.97\figurewidth,
ymin=0, ymax=1.0,
xmin=0, xmax=8e-3,
xlabel={Latency [s]},
ymajorgrids,
]

\addplot [semithick, color0, mark=*, mark size=2, mark repeat=10, mark options={fill=white}] table {./figures/worker_cdf/data/azure/cdf_communication_39_1.csv};
\addlegendentry{Worker $1$ (comm.)}

\addplot [semithick, color1, densely dashed, mark=*, mark size=2, mark repeat=10, mark options={fill=white, solid}] table {./figures/worker_cdf/data/azure/cdf_communication_39_11.csv};
\addlegendentry{Worker $2$ (comm.)}

\addplot [semithick, color0, mark=triangle*, mark size=2.5, mark repeat=10, mark options={fill=white}] table {./figures/worker_cdf/data/azure/cdf_compute_39_1.csv};
\addlegendentry{Worker $1$ (comp.)}

\addplot [semithick, color1, densely dashed, mark=triangle*, mark size=2.5, mark repeat=10, mark options={fill=white, solid}] table {./figures/worker_cdf/data/azure/cdf_compute_39_11.csv};
\addlegendentry{Worker $2$ (comp.)}

\addplot [semithick, black, dashed] table {./figures/worker_cdf/data/azure/cdf_fit_communication_39_1.csv};
\addlegendentry{Fitted gamma dist.}
\addplot [semithick, black, dashed, forget plot] table {./figures/worker_cdf/data/azure/cdf_fit_communication_39_11.csv};
\addplot [semithick, black, dashed, forget plot] table {./figures/worker_cdf/data/azure/cdf_fit_compute_39_1.csv};
\addplot [semithick, black, dashed, forget plot] table {./figures/worker_cdf/data/azure/cdf_fit_compute_39_11.csv};

\end{axis}%

\end{tikzpicture}}
  \caption{Empirical CDF of the per-iteration latency of the two workers in \cref{fig:timeseries}. Worker $2$ is, on average, $14$\% slower than worker $1$. Black dashed lines indicate fitted gamma distributions.}
  \label{fig:worker_cdf}
  \vspace{-5mm}
\end{figure}
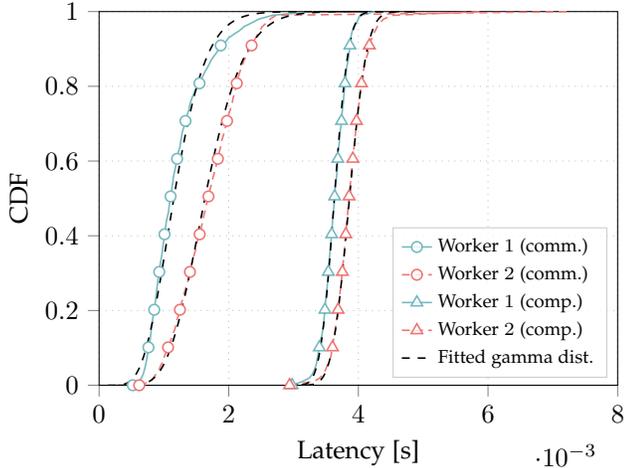%

We find that the latency distribution of workers may change significantly over time, but that these changes typically occur quickly and that the distribution remains approximately constant between changes. Here, we characterize the latency of individual workers while in steady state. Our results are based on traces collected from running many iterations of \cref{eq:kernel} in sequence over a set of workers. For each iteration, we wait for all workers to return their result before proceeding to the next iteration. For this section, we have deliberately chosen traces for which the latency distribution does not change significantly throughout the computation.

In \cref{fig:timeseries}, we plot the communication latency (circles) and computation  latency (triangles) recorded for two workers over \num{100} iterations (out of a total of \num{1600}) on Azure. 
Note that the average latency differs between the two workers; worker $2$ is about $14$\% slower. We show the associated cumulative distribution functions (CDFs) in \cref{fig:worker_cdf}. Now, for a set of workers, we model the latency of the $i$-th worker by the random variable 
\begin{equation} \label{eq:model}
  X_i^{(b, c)} = Y_i^{(b)} + Z_i^{(c)},
\end{equation}
where $Y_i^{(b)}$ and $Z_i^{(c)}$ are random variables associated with the communication and computation latency, respectively, of the worker, when the number of bytes communicated is $b$ and the computational load is $c$.\footnote{The model proposed in~\cite{Yang2019}, where the latency of each worker is assumed to take on one of two discrete values, is similar to ours in the sense that latency may differ between workers. However, for our model, latency takes on values according to a continuous probability distribution.} We often omit the superscripts $b$ and $c$. 

We find that the communication and computation latency of workers on Azure and AWS is well-approximated by independent gamma-distributed random variables,\footnote{In several previous works, latency is modeled by shifted exponential-distributed random variables. These models are related, since the sum of several exponential random variables is gamma-distributed. Hence, a possible interpretation is that the latency we record is the sum of the latency of several smaller computations, each of which has exponentially distributed latency.} but that the parameters of these distributions typically differ between workers, i.e., probability distributions have to be fitted to the particular set of workers used for each computation, especially for systems like Azure CycleCloud and AWS ParallelCluster, which create new worker instances on-demand at the start of a computation. Failing to account for these differences can significantly reduce the accuracy of predictions made using the model; see~\cref{sec:orderstats,fig:orderstats}.

\subsection{Variability over time}
\label{sec:statetransition}

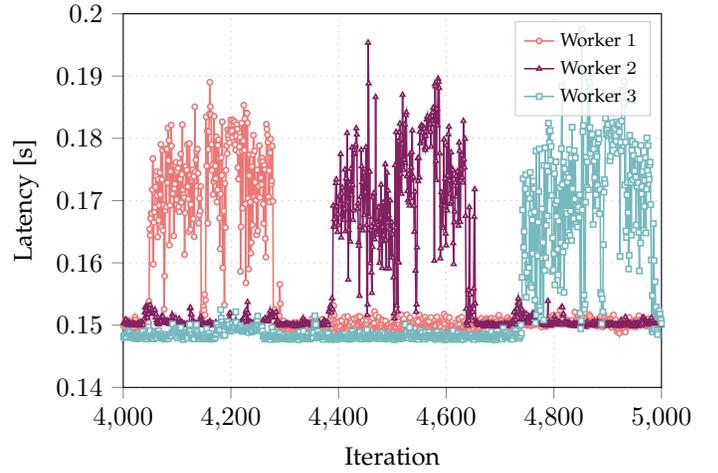
\begin{figure}[t]
\begin{tikzpicture}

\definecolor{color0}{HTML}{75b9be}
\definecolor{color1}{HTML}{ee7674}
\definecolor{color2}{HTML}{832161}
\definecolor{color3}{HTML}{eab464}
\definecolor{color4}{HTML}{335145}

\pgfplotsset{
legend image code/.code={
\draw[mark repeat=2,mark phase=2]
plot coordinates {
(0cm,0cm)
(0.2cm,0cm)        
(0.4cm,0cm)         
};%
}
}

\begin{axis}[
name=plot1,
legend cell align={left},
legend style={font=\scriptsize, fill opacity=0.8, draw opacity=1, text opacity=1, at={(0.98,0.98)}, anchor=north east, draw=white!80!black},
tick align=outside,
tick pos=left,
grid style={gray,opacity=0.5,dotted},
xlabel={Iteration},
xmajorgrids,
height=\figureheight,
width=\figurewidth,
xmin=4000, xmax=5000,
ylabel={Latency [s]},
ymajorgrids,
ymin=0.14, ymax=0.2,
ytick distance=1e-2,
]

\addplot [semithick, color1, mark=*, mark size=1, mark options={fill=white,solid}] 
table {./figures/timeseries_burst/data/aws/timeseries_compute_664_6.csv};
\addlegendentry{Worker $1$}
\addplot [semithick, color2, mark=triangle*, mark size=1, mark options={fill=white,solid}] 
table {./figures/timeseries_burst/data/aws/timeseries_compute_664_7.csv};
\addlegendentry{Worker $2$}
\addplot [semithick, color0, mark=square*, mark size=1, mark options={fill=white,solid}] 
table {./figures/timeseries_burst/data/aws/timeseries_compute_664_1.csv};
\addlegendentry{Worker $3$}

\end{axis}

\end{tikzpicture}
  \caption{Per-iteration computation latency of $3$ workers (out of $\Nn=36$) on AWS, with computational load $c=\num{7.565560560000001e7}$. Workers typically experience bursts of high latency.}
  \label{fig:timeseries_burst}  
  \vspace{-5mm}
\end{figure}

The latency distribution of any particular worker typically changes over time. In particular, as a consequence of the design of cloud computing systems, where multiple virtual machines share the same physical host machine, workers experience \emph{bursts} of higher latency. For example, performing memory-intensive operations, such as matrix multiplication, can more than halve the bandwidth available to other threads on the same machine~\cite{Langguth2018}.\footnote{This is known as the \emph{noisy neighbor} problem.} Further, computations managed by cluster schedulers, such as Borg or Kubernetes, are often only guaranteed a very low fraction of the CPU cycles of the server it is assigned to, but may opportunistically use any cycles not used by other computations~\cite{Verma2015}, \cite[Ch.~14.3]{Luksa2017}, potentially resulting in large performance fluctuations.

In~\cref{fig:timeseries_burst}, we show an example of such high-latency bursts, with the average latency of each of $3$ workers out of the $\Nn=36$ workers used for a particular computation on AWS increasing by about $12$\% for about one minute.\footnote{Similar behavior was observed on AWS in~\cite{Maji2015,Javadi2017}.} The entire computation lasts for about $30$ minutes, and most of the $36$ workers experience at least one such burst over this time. Further, at least one worker is currently experiencing a burst of high latency for about $40$\% of the iterations. This problem becomes more severe for a larger number of workers\textemdash for computations consisting of hundreds of workers, the probability that no worker is currently experiencing a latency burst is close to zero.

\section{Predicting the latency of distributed gradient computations}
\label{sec:latency}

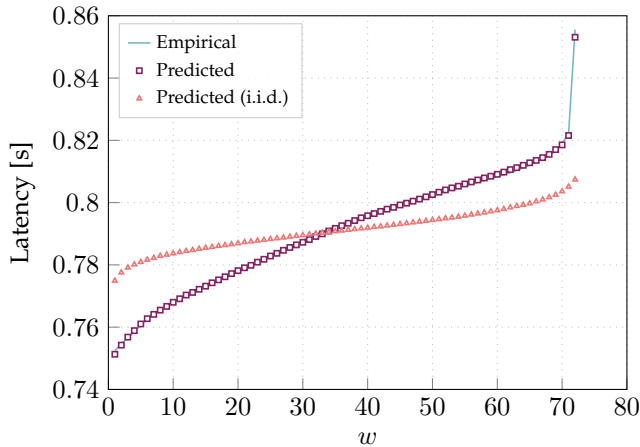
\begin{figure}[t]
\begin{tikzpicture}

\definecolor{color0}{HTML}{75b9be}
\definecolor{color1}{HTML}{ee7674}
\definecolor{color2}{HTML}{832161}
\definecolor{color3}{HTML}{eab464}
\definecolor{color4}{HTML}{335145}

\begin{axis}[
legend cell align={left},
legend style={font=\scriptsize, fill opacity=0.8, draw opacity=1, text opacity=1, at={(0.02,0.98)}, anchor=north west, draw=white!80!black},
tick pos=left,
grid style={gray,opacity=0.5,dotted},
xlabel style={align=center}, xlabel={$w$},
xtick distance=10,
xmajorgrids,
height=\figureheight,
width=0.97\figurewidth,
xmin=0, xmax=80,
ylabel={Latency [s]},
ymajorgrids,
ymin=0.74, ymax=0.86,
ytick distance=0.02,
]

\addplot [semithick, color0] 
table {./figures/orderstats/data/azure/orderstats_2.csv};
\addlegendentry{Empirical}

\addplot [semithick, dashed, only marks, color2, mark=square*, mark size=1, mark options={fill=white,solid}] 
table {./figures/orderstats/data/azure/orderstats_gamma_gamma_2.csv};
\addlegendentry{Predicted}

\addplot [semithick, dashed, only marks, color1, mark=triangle*, mark size=1, mark options={fill=white,solid}] 
table {./figures/orderstats/data/azure/orderstats_global_gamma_2.csv};
\addlegendentry{Predicted (i.i.d.)}





\end{axis}

\end{tikzpicture}
  \caption{Average latency of the $w$-th fastest worker out of $\Nn=72$ workers for a computation 
  on Azure, and predicted latency, where the per-worker latency is modeled as either independent, but not necessarily identically distributed, or i.i.d., between workers. The i.i.d. assumption can significantly reduce accuracy.}
  \label{fig:orderstats}
  \vspace{-5mm}
\end{figure}

Here, we show how to efficiently estimate the latency of the $w$-th fastest worker ($w \leq \Nn$) of a set of workers, i.e., the $w$-th \emph{order statistic} of the per-worker latency. Later, we use these predictions for dynamic load-balancing to minimize latency variations between workers (see~\cref{sec:loadbalancing}). Throughout this section, we have deliberately chosen traces where workers are operating in steady state. When used for load-balancing, we account for bursts by dynamically updating the estimate of the latency distribution associated with each worker. We first consider the case where all workers are available at the start of each iteration (\cref{sec:orderstats}), before considering the more realistic case where workers may remain unavailable over several iterations (\cref{sec:iterative}).

\subsection{Order statistics latency}
\label{sec:orderstats}

From the distributions of $Y_i$ and $Z_i$ for each worker, we can compute the distribution of the latency of the $w$-th fastest worker. However, the computational complexity of doing so analytically may be prohibitively high when the number of workers is large. Instead, we use Monte Carlo integration. The complexity of sampling from the latency of the $w$-th fastest worker is linear in the number of workers, since we can first sample from $Y_i$ and $Z_i$ for each worker and then find the $w$-th smallest value of the resulting list in linear time, e.g., using the Quickselect algorithm. Through this process we can estimate, e.g., the expected latency of the $w$-th fastest worker. 

In \cref{fig:orderstats}, we plot the average latency of the $w$-th fastest worker out of $\Nn=72$ workers for a particular computation with $b=30048$ and $c=\num{4.5452629936797535e8}$ on Azure. We also plot predictions made using Monte Carlo integration as explained above, and, for reference, predictions made by the commonly adopted i.i.d.\ model, where the latency of each worker is modeled by a random variable with mean and variance equal to the global mean and variance computed across all workers.\footnote{We model the latency distribution by a gamma distribution, which we find provides more accurate predictions than the more commonly used shifted exponential distribution.} The proposed model yields an accurate prediction of the empirical performance, while assuming that latency is i.i.d. between workers can significantly reduce accuracy.

\subsection{Order statistics latency of iterative computations}
\label{sec:iterative}

In~\cref{sec:orderstats}, we considered order statistics in cases where all workers are available at the start of each iteration. However, for straggler-resilient methods, we wish to proceed to the next iteration immediately after receiving results from the $w$ fastest workers, without waiting for the remaining $\Nn - w$ workers, which may remain unavailable over several subsequent iterations.

Here, we show how to estimate latency in this scenario. Denote by $T_w^{(t)}$ the time at which the $t$-th iteration of an iterative computation, for which the coordinator waits for the $w$-th fastest worker in each iteration, is completed (i.e., the latency of the $t$-th iteration is $T_w^{(t)} - T_w^{(t-1)}$). We wish to simulate the time series process $T_w^{(1)}, \dots, T_w^{(\ell)}$, where $\ell$ is the number of iterations. We do so by using a two-state model, where workers are either idle or busy. First, each worker has a local first-in-last-out task queue of length \num{1}. If the $i$-th worker is idle and there is a task in its queue, it immediately removes the task from the queue and becomes busy for a random amount of time, which is captured by the random variable $X_i$ (recall that we can sample from $X_i$, see \cref{sec:steadystate}), before becoming idle again. At the start of each iteration, the coordinator assigns a task to each worker, and once $w$ of those tasks have been completed, the coordinator proceeds to the next iteration.

Using this model, we can efficiently simulate realizations of $T_w^{(1)}, \dots, T_w^{(\ell)}$ by using a priority queue data structure (see, e.g., \cite{Boas1976}) to map the index of each worker to the next time at which it will transition from busy to idle. This strategy is typically referred to as \emph{event-driven} simulation. By performing such simulations we can estimate, e.g., the expected time required to perform $\ell$ iterations, in a manner that accounts for the fact that workers may remain unavailable over several iterations. We provide an implementation of such a simulator in~\cite{dsagsource}.

\begin{figure}[t]
  \centering
\begin{tikzpicture}

\definecolor{color0}{HTML}{75b9be}
\definecolor{color1}{HTML}{ee7674}
\definecolor{color2}{HTML}{832161}
\definecolor{color3}{HTML}{eab464}
\definecolor{color4}{HTML}{335145}

\begin{axis}[
name=plot1,
legend cell align={left},
legend style={font=\scriptsize, fill opacity=0.8, draw opacity=1, text opacity=1, at={(axis cs: 1,9.8)}, anchor=north west, draw=white!80!black},
tick align=outside,
tick pos=left,
grid style={gray,opacity=0.5,dotted},
xlabel={Iteration},
xmajorgrids,
height=\figureheight,
width=0.97\figurewidth,
xmin=0, xmax=100,
ylabel={Cumulative latency [s]},
ymajorgrids,
ymin=0, ymax=10,
]

\addplot [semithick, color0] 
table {./figures/cumulative_timeseries/data/cumulative_time_773.csv};
\addlegendentry{Empiric}

\addplot [semithick, color0, mark=triangle*, mark size=1, only marks, mark repeat=5, mark options={fill=white}] 
table {./figures/cumulative_timeseries/data/gamma/cumulative_time_sim_773.csv};
\addlegendentry{Predicted}

\addplot [semithick, color0, mark=*, mark size=1, only marks, mark repeat=5, mark options={fill=white}] 
table {./figures/cumulative_timeseries/data/gamma/cumulative_time_simint_773.csv};
\addlegendentry{Predicted, event-driven}

\addplot [semithick, color1, dashed] 
table {./figures/cumulative_timeseries/data/cumulative_time_769.csv};
\addlegendentry{Empiric}

\addplot [semithick, color1, mark=triangle*, mark size=1, only marks, mark repeat=5, mark options={fill=white}] 
table {./figures/cumulative_timeseries/data/gamma/cumulative_time_sim_769.csv};
\addlegendentry{Predicted}

\addplot [semithick, color1, mark=*, mark size=1, only marks, mark repeat=5, mark options={fill=white}] 
table {./figures/cumulative_timeseries/data/gamma/cumulative_time_simint_769.csv};
\addlegendentry{Predicted, event-driven}




\end{axis}

\end{tikzpicture}
  \caption{Cumulative latency over \num{100} iterations. Blue curves correspond to $w=9$ and red curves to $w=72$. Each iteration ends once $w$ workers have completed their task. For $w < \Nn$, we need to account for the case where workers remain unavailable over several iterations, which we do using the model based on event-driven simulations.}
  \label{fig:cumulative_time}
  \vspace{-5mm}
\end{figure}
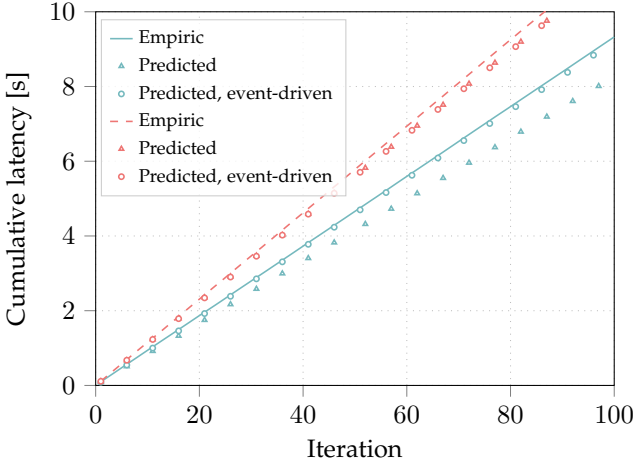%

In \cref{fig:cumulative_time}, we plot the cumulative latency over $100$ iterations for two jobs, with $b=\num{30048}$, $c=\num{7.5754383227995895e6}$, and $\Nn=72$ on AWS, where, in one job, we wait for $w=9$ workers (blue curves) and, in the other, for all $w=\Nn=72$ workers (red curves). We also plot the predictions made by the proposed model based on event-driven simulations, which accounts for the interaction between iterations, and the model described in \cref{sec:orderstats}, which does not. For $w=\Nn=72$, both models give  accurate predictions. However, for $w=9 < \Nn$, the model of \cref{sec:orderstats} underestimates the overall latency, since it does not account for the case where workers remain unavailable over multiple iterations. The model based on event-driven simulations remains accurate. 

\section{DSAG}
\label{sec:dsag}

In this section, we consider learning in cloud computing systems. In particular, we want an optimization method that i) is able to make progress even when some workers fail to respond, ii) has fast initial convergence, similar to SGD, which is achieved by performing many fast, but inexact, iterations, iii) eventually converges to the optimum, iv) allows for dynamic load-balancing, and v) has low update complexity. GD and SAG fail points i) and iv), SGD fails point iii), and coded computing methods fail either point ii) or iii), and, in most cases, points iv) and v).

To address i)--v), we introduce DSAG, which adapts SAG to distributed environments with heterogenous and straggling workers. As with SAG, the key idea is to cache stale subgradients. However, unlike SAG, DSAG utilizes subgradients computed in previous iterations that arrive late. Further, DSAG allows for load-balancing by dynamically changing the number of data partitions (and hence the number of samples that make up each partition). DSAG meets all of the above criteria.

DSAG works as follows. Denote by
\begin{equation} \notag
  \bY_{i:j}^{(t)} \define \sum_{k=i}^j \nabla f_k\left(\bV^{(t)}\right)
\end{equation}
the subgradient computed from samples $i$ through $j$, where $j \geq i$. The coordinator maintains a set of such subgradients, denoted by $\mathcal{Y}$, which we refer to as the gradient cache. Upon receiving a subgradient $\bY_{i:j}^{(t)}$ from a worker, the coordinator first selects the subset of overlapping subgradients
\begin{equation} \notag
  \mathcal{Y}' \define
  \left\{ 
    \bY_{{i'}:{j'}}^{(t')} \in \mathcal{Y} 
    \;:\;
    i \leq i' \leq j \text{ or } i \leq j' \leq j 
  \right\}.
\end{equation}
If any such subgradient is more recent than the received subgradient (i.e., if $t' \geq t$ for some $\bY_{{i'}:{j'}}^{(t')} \in \mathcal{Y}'$), the process is aborted and the received subgradient discarded. Otherwise, the overlapping subgradients are discarded in favour of the received subgradient, i.e.,
\begin{equation} \notag
  \mathcal{Y} \leftarrow \left( \mathcal{Y} \setminus \mathcal{Y}' \right)
  \cup \left\{ \bY_{i:j}^{(t)} \right\}.
\end{equation}
This process allows for changing partition boundaries at runtime, e.g., due to load-balancing, and can be implemented efficiently by storing the elements of $\mathcal{Y}$ as nodes in a tree data structure.\footnote{When using a tree data structure, the complexity of deleting and inserting subgradients is in $\mathcal{O}\left(\log \left\vert \mathcal{Y} \right\vert \right)$.} Denote by
\begin{equation} \label{eq:sagH}
  \bH \define \sum_{y\in \mathcal{Y}} y
\end{equation}
the sum of the elements of $\mathcal{Y}$. The coordinator maintains this sum by assigning
\begin{equation} \notag
  \bH \leftarrow \bH + \bY_{i:j}^{(t)} - \sum_{y \in \mathcal{Y}'} y
\end{equation}
whenever a subgradient $\bY_{i:j}^{(t)}$ is inserted into $\mathcal{Y}$. Finally, $\bH$ is used in place of the exact gradient $\nabla F$ to update $\bV^{(t)}$, i.e.,
\begin{equation} \label{eq:saggd}
  \bV^{(t+1)} = G\left( \bV^{(t)} - \eta \left( \frac{1}{\xi} \bH + \nabla R\left(\bV^{(t)}\right) \right) \right),
\end{equation}
where $\xi$ is the fraction of samples covered by the elements of $\mathcal{Y}$. Scaling the gradient in this way improves the rate of convergence for the iterations before the coordinator has received subgradients covering all samples.\footnote{A similar scaling is used by SAG.}

We remark that if there exists $\bY_{i':j'}^{(t')}$ in $\mathcal{Y}$ such that $i' = i$ and $j' = j$, the existing element can be updated in-place. In this case, and if the received subgradient is computed from the most recent iterate, the update process degrades to that of SAG.

\subsection{Distributed implementation}
\label{sec:dsag_dist}

Here, we describe our distributed implementation of DSAG. In particular, we wish to maintain predictable and low latency in the presence of stragglers. For SAG or SGD, this can be achieved by only waiting for a subset of workers to return in each iteration, and ignoring any results computed by straggling workers. However, since the same workers are likely to be stragglers for extended periods of time, the subgradients received from the fastest subset of workers by the coordinator will not be selected uniformly at random, unless all workers store the entire dataset or the coordinator waits for all workers. This can significantly reduce the rate of convergence, since parts of the dataset may never factor into the learning process (see~\cref{sec:convergence,fig:convergence}).

DSAG addresses this shortcoming by utilizing stale results and through dynamic load-balancing. In particular, at the $t$-th iteration, the coordinator waits until it has received subgradients computed from $\bV^{(t)}$ from at least $w$ workers. During this time, the coordinator may also have received subgradients from previous iterations, which the coordinator stores if they are less stale than the currently stored subgradients it would replace. Further, we allow for a small margin, such that after receiving the $w$-th fresh subgradient, the coordinator waits for \num{2}\% longer\textemdash collecting any subgradients received during this time\textemdash before updating the iterate. We find that doing so can improve the rate of convergence at the expense of a small increase in latency, especially when combined with load-balancing. We explain the load-balancing strategy that we propose in~\cref{sec:loadbalancing}.

\subsection{Convergence of DSAG}

DSAG builds upon the SAG method, for which the error
\begin{equation} \notag
  F\left(\bV^{(t)}\right) - F\left(\bV^*\right),
\end{equation}
where $\bV^*$ is the optimum, decreases with $\mathcal{O}(1/t)$ and $\mathcal{O}(\rho^t)$, for some $\rho < 1$, for convex and strongly convex problems, respectively~\cite{Schmidt2017}. We do not have convergence proofs for DSAG\textemdash the analysis of asynchronous optimization methods is notoriously challenging, and the analysis of SAG is already complex\textemdash but we make a few remarks to relate the behavior of DSAG to that of SAG.

SAG updates one subgradient, selected uniformly at random over all partitions, at each iteration, and does not make use of stale subgradients. DSAG differs by updating one or more subgradients per iteration, and in that some of the updated subgradients may have been computed from a previous iterate, provided they are less stale than the replaced subgradients. Hence, the subgradients utilized by DSAG are at least as fresh as those used by SAG. Second, DSAG, unlike SAG, may discard cached subgradients if it receives a subgradient that is not aligned with an already cached subgradient (SAG does not support changing the partition boundaries at runtime).

Hence, we conjecture that the rate of convergence of DSAG is at least as good as that of SAG for iterations when no subgradients are discarded, and that it is worse than that of SAG in iterations where cached subgradients have been discarded, and until the discarded entries have been repopulated. We present empirical results that support this conjecture, see~\cref{sec:convergence}.

\section{Load-balancing}
\label{sec:loadbalancing}

\begin{figure}[t]
  \centering
  \begin{tikzpicture}

    \definecolor{color0}{HTML}{75b9be}
    \definecolor{color1}{HTML}{ee7674}
    \definecolor{color2}{HTML}{832161}
    \definecolor{color3}{HTML}{eab464}
    \definecolor{color4}{HTML}{335145}
    
    \begin{axis}[
    name=plot1,
    legend cell align={left},
    legend style={font=\scriptsize, fill opacity=0.8, draw opacity=1, text opacity=1, at={(0.98,0.98)}, anchor=north east, draw=white!80!black},
    tick pos=left,
    grid style={gray,opacity=0.5,dotted},
    xmajorgrids,
    xticklabels=\empty,
    height=\figureheight,
    width=\figurewidth,
    xmin=0, xmax=140,
    ylabel={Latency [s]},
    ymajorgrids,
    ymin=0.2, ymax=0.8,
    ytick distance=0.1,
    ]


    \addplot [semithick, color2]
    table {./figures/loadbalancing/data/balanced/timeseries_6_7.csv};
    \addlegendentry{Regular workers}    

    \addplot [semithick, color0]
    table {./figures/loadbalancing/data/balanced/timeseries_6_4.csv};
    \addlegendentry{Slow workers}    
    
    \addplot [semithick, color1]
    table {./figures/loadbalancing/data/balanced/timeseries_6_1.csv};
    \addlegendentry{Fast workers}

    \addplot [semithick, color1, forget plot]    
    table {./figures/loadbalancing/data/balanced/timeseries_6_2.csv};

    \addplot [semithick, color1, forget plot]
    table {./figures/loadbalancing/data/balanced/timeseries_6_3.csv};

    \addplot [semithick, color0, forget plot]
    table {./figures/loadbalancing/data/balanced/timeseries_6_5.csv};
    
    \addplot [semithick, color0, forget plot]
    table {./figures/loadbalancing/data/balanced/timeseries_6_6.csv};    

    \addplot [semithick, color2, forget plot]
    table {./figures/loadbalancing/data/balanced/timeseries_6_8.csv};

    \addplot [semithick, dashed, black]
    table {./figures/loadbalancing/data/balanced/timeseries_overall_6.csv};
    \addlegendentry{Overall latency}

    \addplot [thin, gray]
    table {%
    12 0
    12 1
    12 0
    55 0
    55 1
    55 0
    59 0
    59 1
    59 0
    63 0
    63 1
    63 0
    72 0
    72 1
    72 0
    76 0
    76 1
    76 0
    100 0
    100 1
    100 0
    104 0
    104 1
    104 0
    108 0
    108 1
    108 0
    112 0
    112 1
    112 0
    117 0
    117 1
    117 0
    128 0
    128 1
    128 0
    134 0
    134 1
    134 0
    151 0
    151 1
    151 0
    };

\end{axis}

\begin{axis}[
    name=plot2,
    at=(plot1.below south west), anchor=above north west,
    legend cell align={left},
    legend style={font=\scriptsize, fill opacity=0.8, draw opacity=1, text opacity=1, at={(0.98,0.98)}, anchor=north east, draw=white!80!black},
    tick pos=left,
    grid style={gray,opacity=0.5,dotted},
    xlabel style={align=center}, xlabel={Iteration},
    xmajorgrids,
    height=\figureheight,
    width=\figurewidth,
    xmin=0, xmax=140,
    ylabel={Latency [s]},
    ymajorgrids,
    ymin=0.2, ymax=0.8,
    ytick distance=0.1,    
    ]    
    \addplot [semithick, color2]
    table {./figures/loadbalancing/data/unbalanced/timeseries_3_7.csv};

    \addplot [semithick, color0]
    table {./figures/loadbalancing/data/unbalanced/timeseries_3_4.csv};
    
    \addplot [semithick, color1]
    table {./figures/loadbalancing/data/unbalanced/timeseries_3_1.csv};

    \addplot [semithick, color1]    
    table {./figures/loadbalancing/data/unbalanced/timeseries_3_2.csv};

    \addplot [semithick, color1]
    table {./figures/loadbalancing/data/unbalanced/timeseries_3_3.csv};

    \addplot [semithick, color0]
    table {./figures/loadbalancing/data/unbalanced/timeseries_3_5.csv};
    
    \addplot [semithick, color0]
    table {./figures/loadbalancing/data/unbalanced/timeseries_3_6.csv};    

    \addplot [semithick, color2]
    table {./figures/loadbalancing/data/unbalanced/timeseries_3_8.csv};

    \addplot [semithick, dashed, black]
    table {./figures/loadbalancing/data/unbalanced/timeseries_overall_3.csv};    

\end{axis}            
\end{tikzpicture}
  \caption{Per-worker latency for $\Nn = 8$ workers with (top) and without (bottom) load-balancing, when waiting for all workers (i.e., $w = \Nn$). We artificially slow down \num{3} randomly selected workers (blue lines) after \num{40} iterations, and speed up another set of \num{3} randomly selected workers (red lines) after \num{90} iterations. Note that there is some natural variation in addition. The load-balancer automatically re-balances the workload in the iterations marked with gray lines. For the final $20$ iterations, the overall latency of the unbalanced system is more than twice that of the system with load-balancing.}
  \vskip -5mm  
  \label{fig:loadbalancing}
\end{figure}
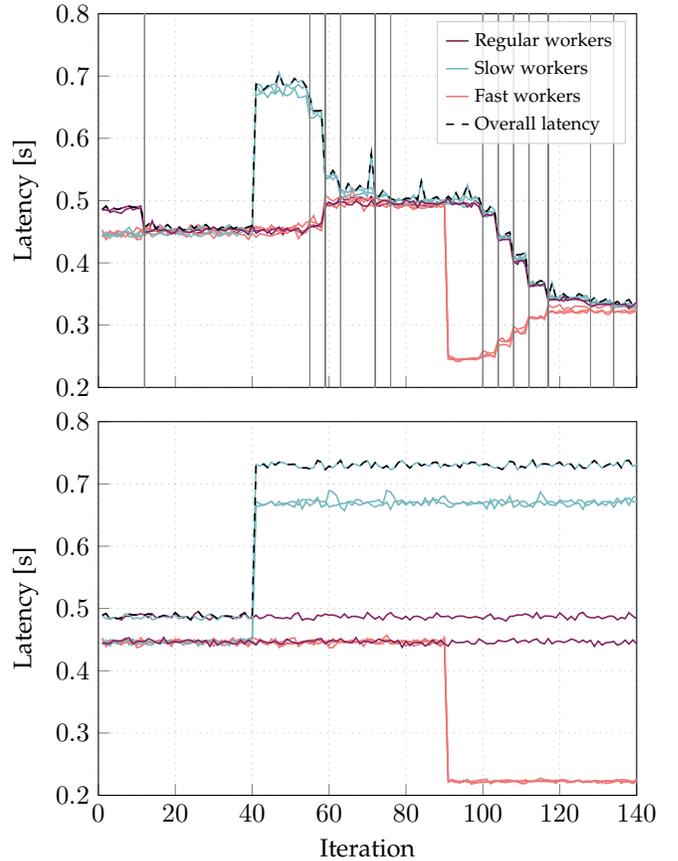

Recall that computing speed typically differs between workers and may change over time (see~\cref{sec:latencymodel}). Unless these differences are accounted for, fast workers typically spend a significant amount of time waiting for slower ones, and some workers may never be among the $w$ fastest ones. Here, we propose a strategy to dynamically adjust the size of the data partitions stored by each worker to alleviate this issue. The process consists of three steps:
\begin{enumerate}
  \item Latency profiling to estimate the probability distribution of $Z_i$ and $Y_i$ for each worker based on recorded latency (see~\cref{sec:profiler}).
  \item Optimizing the number of subpartitions for each worker using simulations based on the latency model of \cref{sec:iterative} to predict the impact of each change (see~\cref{sec:optimizer}).
  \item Re-partitioning the local dataset for any workers for which the number of subpartitions has changed (see~\cref{sec:repartitioning}).
\end{enumerate}
All three steps are performed asynchronously in parallel and are running continuously in the background. In particular, whenever the optimizer finishes, it is restarted to include any new latency recordings.\footnote{The load-balancer proposed in~\cite{Javadi2017} takes a similar approach, but is designed for web services.} We show how load-balancing affects latency in~\cref{fig:loadbalancing}, and we describe the three steps in detail next.

\subsection{Latency profiler}
\label{sec:profiler}

The latency profiler is responsible for estimating the mean and variance of the communication and computation latency of each worker, and for providing these to the optimizer. It takes as its input latency recorded both by the coordinator and the workers themselves. In particular, for each worker, the coordinator records the time between sending an iterate to the worker and receiving a response. Meanwhile, the workers record the time between starting to process a received iterate and having a response ready, and  include this recording in their responses. 

For each worker, we take the latency recorded by the worker as a sample of the computation latency, and the difference between the latency recorded by the worker and coordinator as a sample of the communication latency, i.e., for the $i$-th worker, as realizations of $Z_i$ and $Y_i$, respectively. Hence, we record the round-trip communication latency, which includes the time required for data to be sent over the wire and any queuing at either end.

Next, for each worker, the profiler computes the sample mean and variance over a moving time window, i.e., samples older than a given deadline (in seconds) are discarded before processing. Choosing a window size involves making a trade-off\textemdash a larger window size makes statistics computed over it less noisy, but increases the time needed for the profiler to adapt to changes.\footnote{We use a window size of $10$ seconds, which we find is a good trade-off for the applications we consider.} We denote by $e_{\mathsf{Y}, i}$ and $v_{\mathsf{Y}, i}$ the mean and variance of the communication latency of the $i$-th worker, and by $e_{\mathsf{Z}, i}$ and $v_{\mathsf{Z}, i}$ the mean and variance of the computation latency,  computed as described above. For each worker, whenever new latency recordings are available, the mean and variance of its communication and computation latency are re-computed and sent to the optimizer, which uses them to fit probability distributions.\footnote{The shape and scale parameter of a gamma-distributed random variable with mean $e$ and variance $v$ is $e^2/v$ and $v/e$, respectively.}

\subsection{Optimizer}
\label{sec:optimizer}

For each worker, we tune its workload by changing the number of subpartitions that the data it stores locally is divided into. The optimizer takes as its input the most recent statistics computed by the profiler and a vector $\bm{p} = [p_1, \dots, p_{\Nn}]$ containing the current number of subpartitions for each worker, and returns an updated vector $\bm{p}' = [p_1', \dots, p_{\Nn}']$. For any solution $\bm{p}$, we impose a constraint on the expected overall per-iteration \emph{contribution}, which we define as
\begin{equation} \notag
  h(\bm{p}) \define \sum_{i=1}^{\Nn} h_i(\bm{p}), \text{ with } h_i(\bm{p}) \define \frac{u_i(\bm{p}) n_i}{p_i n},
\end{equation}
where $n_i$ is the number of samples stored by the $i$-th worker and $u_i(\bm{p})$ the fraction of iterations that the $i$-th worker delivers a fresh result in. Hence, $h_i(\bm{p})$ is a measure of the extent to which the $i$-th worker contributes to the learning process. Note that $u_i$ is a nonlinear function of $\bm{p}$, i.e., it depends on the workload of the entire set of workers. The goal of the optimizer is to minimize latency variation between workers within this constraint. More formally, its goal is to solve
\begin{equation} \label{eq:objective}
  \begin{aligned}
    \argmin_{\bm{p}'} \quad &
    \frac{
        \max\left\{e_{\mathsf{X}, 1}', \dots, e_{\mathsf{X}, \Nn}'\right\}
      }{
        \min\left\{e_{\mathsf{X}, 1}', \dots, e_{\mathsf{X}, \Nn}'\right\}
      } \\
    \text{s.t.} \quad & h(\bm{p}') \geq h_\mathsf{min}, \\
  \end{aligned}
\end{equation}
where $h_\mathsf{min}$ is the constraint, and $e_{\mathsf{X}, i}'$ is the expected overall latency of the $i$-th worker if its local dataset is split into $p_i'$ subpartitions. Throughout the optimization process, we use the approximations
\begin{equation} \notag
  e_{\mathsf{Z}, i}' \define e_{\mathsf{Z}, i} \frac{p_i}{p_i'}
  \;,\;\;\;
  v_{\mathsf{Z}, i}' \define v_{\mathsf{Z}, i} \frac{p_i^2}{{p_i'}^2},
\end{equation}
and 
\begin{equation} \notag
  e_{\mathsf{X}, i}' \define e_{\mathsf{Y}, i} + e_{\mathsf{Z}, i}',
\end{equation}
where $p_i$ is the current number of subpartitions of the $i$-th worker. Hence, we linearize the mean and variance of the computation latency around the value of $p_i$ for which it was recorded.\footnote{This linearization is motivated by~\cref{fig:linear_latency}. If latency has been measured for several different values of $p_i$, we use a weighted average over the values of $p_i$ for which we have recordings.}

It is difficult to compute $u_i$, and thus $h$, analytically, but $u_i$ can be estimated via event-driven simulations as explained in~\cref{sec:iterative}.\footnote{With our implementation, for $\Nn = 100$ workers and $w=50$, simulating $100$ iterations of the learning process takes about $1.5$ milliseconds.} However, this requires that the optimizer i) is robust against noise in the estimates of $u_i$, and ii) evaluates $u_i$ a small enough number of times to be computationally fast enough  to provide useful solutions in time. We find that traditional optimization techniques that, e.g., rely on gradients, fail the first criteria, while \emph{meta-heuristic} techniques (e.g., evolutionary algorithms) fail the second. Hence, we propose an optimizer that solves~\cref{eq:objective} by making small changes to $\bm{p}$ in an iterative fashion.

\begin{algorithm}[!t]
  \caption{Load-balancer}
  \label{al:optimizer}  
  \begin{algorithmic}
    \Procedure{Optimize}{$\bm{p}$}
      \State $\bm{p}' \gets \bm{p}$
      \State $i \gets \argmax\left[ e_{\mathsf{X}, 1}', \dots, e_{\mathsf{X}, \Nn}' \right]$ \Comment{Slowest worker}
      \For{$j = 1, \dots, \Nn$}
        \State $p_j' \gets \left\lfloor \frac{e_{\mathsf{Z}, j} p_j}{e_{\mathsf{Y}, i} + e_{\mathsf{Z}, i} - e_{\mathsf{Y}, j}} \right\rfloor$ \Comment{Equalize total latency}
      \EndFor
      \While{$h(\bm{p}') < h_\mathsf{min}$}
        \State $i \gets \argmin\left[ e_{\mathsf{X}, 1}', \dots, e_{\mathsf{X}, \Nn}' \right]$ \Comment{Fastest worker}
        \State $p_i' \gets \lfloor \num{0.99} \cdot p_i' \rfloor$ \Comment{Increase workload}
      \EndWhile
      \While{$h(\bm{p}') \geq 0.99 \cdot h_\mathsf{min}$}
        \State $i \gets \argmax\left[ e_{\mathsf{X}, 1}', \dots, e_{\mathsf{X}, \Nn}' \right]$ \Comment{Slowest worker}
        \State $p_i' \gets \lceil \num{1.01} \cdot p_i' \rceil$ \Comment{Decrease workload}
      \EndWhile    
      \State \Return{$\bm{p}'$}
    \EndProcedure
    \Loop \Comment{Optimizer main loop}
      \State Collect updated latency statistics from the profiler
      \State $\bm{p} \gets \textsc{Optimize}(\bm{p})$
      \State Distribute the updated vector $\bm{p}$
    \EndLoop
  \end{algorithmic}
\end{algorithm}

At a high level, the optimizer attempts to increase the contribution of workers that are always among the $w$ fastest by giving them more work, without increasing the overall latency. This increases the overall per-iteration contribution, thus giving the optimizer leeway to reduce the overall iteration latency by reducing the workload of the slowest workers. The proposed algorithm is given in \cref{al:optimizer}. Since $h$ is estimated via simulations, we evaluate the constraint with a $1$\% tolerance. Finally, we set the constraint to be
\begin{equation} \notag
  h_\mathsf{min} = h(\bm{p}_\mathsf{0}),
\end{equation}
where $\bm{p}_\mathsf{0}$ is the baseline number of subpartitions for each worker used at the start of the first iteration. This is to ensure that load-balancing does not reduce the rate of convergence.

\subsection{Re-partitioning}
\label{sec:repartitioning}

Whenever the optimizer produces an updated number of subpartitions for a particular worker, the update is included with the next iterate sent to the worker, which re-partitions its local dataset. However, re-partitioning carries a cost, since it invalidates subgradients cached by the coordinator. Here, we show how to minimize the number and impact of such \emph{cache evictions} resulting from re-partitioning. First, we partition the data matrix such that the $i$-th worker stores locally the submatrix
\begin{equation} \notag
  \bX^{(i)} \triangleq \bX_{p_\mathsf{start}(n, \Nn, i):p_\mathsf{stop}(n, \Nn, i)},
\end{equation}
where
\begin{equation} \notag
  p_\mathsf{start}(n, p, i) = \left\lfloor \frac{\left(i - 1\right)n}{p} \right\rfloor + 1
\end{equation}
and
\begin{equation} \notag  
  p_\mathsf{stop}(n, p, i) = \left\lfloor \frac{i n}{p} \right\rfloor,
\end{equation}
with $1 \leq p \leq n$ and $1 \leq i \leq p$. Next, for each worker, we subpartition the data it stores locally, such that, in each iteration, the $i$-th worker processes the matrix
\begin{equation} \notag
  \bX^{(i)}_{p_\mathsf{start}(n_i, p_i, k_i):p_\mathsf{stop}(n_i, p_i, k_i)},
\end{equation}
for some index $k_i$. Hence, we may tune the workload of a worker by sending it a new value $p_i$, which changes the number of samples processed per iteration. The following example shows how doing so leads to cache evictions.
\begin{example}[Re-partitioning] \label{ex:repartitioning}
  Consider a scenario with \num{2} workers, $n_1=n_2=10$ (i.e., $n=20$), and $p_1 = p_2 = 2$, such that the partitions on the first worker are $\bX_{1:5}$ and $\bX_{6:10}$, and $\bX_{11:15}$ and $\bX_{16:20}$ on the second. Now, say that we let $p_1 \leftarrow 3$, such that the partitions on the first worker are $\bX_{1:3}$, $\bX_{4:6}$, and $\bX_{7:10}$. Prior to this change, the coordinator stores gradients corresponding to partitions $\bX_{1:5}$ and $\bX_{6:10}$. Now, if in the next iteration the worker sends to the coordinator the subgradient computed over $\bX_{4:6}$, both of the existing entries need to be evicted before inserting the new subgradient, leading to a lower rate of convergence until the missing cache entries have been populated.
\end{example}

We find that cache evictions due to re-partitioning can significantly reduce the rate of convergence, since the gradient used by DSAG no longer covers all samples of the dataset. We use two strategies to reduce the severity of this issue. First, we refrain from distributing an update $\bm{p}'$ to the workers until doing so would improve the objective function~\cref{eq:objective} by more than some threshold (e.g., $10$\%). Second, we process subpartitions in order to minimize the number of iterations for which evicted cache entries remain empty. More formally, the $i$-th worker stores a counter $k_i$ that it increments in a cyclic fashion each time it receives an iterate, i.e.,\footnote{Note that, when $w < \Nn$, workers, unlike the coordinator, are unaware of the current iteration index since they may have remained unavailable for an arbitrary amount of time.}
\begin{equation} \label{eq:index_update}
  k_i \leftarrow \mathrm{mod}\left(k_i, p_i \right) + 1.
\end{equation}
Next, it computes the gradient with respect to the $k_i$-th of its locally stored partitions. We show the benefit of this approach with the following example.
\begin{example}[Continuation of~\cref{ex:repartitioning}] \label{ex:repartitioning2}
  Immediately after re-partitioning, the coordinator stores subgradients computed over partitions $\bX_{1:5}$ and $\bX_{6:10}$ (we omit partitions stored by the second worker). To minimize cache evictions, over the following $3$ iterations, the first worker sends to the coordinator:
  \begin{enumerate}
    \item The gradient over $\bX_{1:3}$, evicting the gradient over $\bX_{1:5}$, resulting in a cache with the gradients over $\bX_{1:3}$ and $\bX_{6:10}$, leaving the gradient over $\bX_{4:5}$ missing.
    \item The gradient over $\bX_{4:6}$ , evicting the gradient over $\bX_{6:10}$, resulting in a cache with the gradients over $\bX_{1:3}$ and $\bX_{4:6}$, leaving the gradient over $\bX_{7:10}$ missing.
    \item The gradient over $\bX_{7:10}$, resulting in a cache with the gradients over $\bX_{1:3}$, $\bX_{4:6}$, and $\bX_{7:10}$, leaving no missing entries.
  \end{enumerate}
  In this case, the gradients over $\bX_{4:5}$ and $\bX_{7:10}$ are missing from the cache for $1$ iteration each. If instead the worker had started by sending the gradient over $\bX_{4:6}$, either the gradient over $\bX_{1:3}$ or $\bX_{7:10}$ would have been missing for $2$ iterations, and the other for $1$ iteration, resulting in a lower rate of convergence.
\end{example}

\begin{algorithm}[!t]
  \caption{Partition alignment}
  \label{al:alignment}
  \begin{algorithmic}[1]
    \State $k_i \gets \mathrm{mod}\left(k_i, p_i \right) + 1$ \label{alg:line:1}
    \State $k_i' \gets p_\mathsf{trans}(n_i, p_i, p_i', k_i)$    \label{alg:line:2}
    \While{$p_\mathsf{start}(n_i, p_i', k_i') \neq p_\mathsf{start}(n_i, p_i, k_i)$} \label{alg:line:3}
      \State $k_i' \gets k_i' - 1$ \label{alg:line:4}
      \State $k_i \gets p_\mathsf{trans}(n, p_i', p_i, k_i')$ \label{alg:line:5}
    \EndWhile     
    \State $p_i \gets p_i'$ \label{alg:line:7}
    \State $k_i \gets k_i'$ \label{alg:line:8}
  \end{algorithmic}
\end{algorithm}

This approach is most effective if the first sample of the partition processed immediately after a re-partitioning is aligned with the first sample of a partition already in the cache, since otherwise the evicted entries are not re-populated until after a full pass over the data (this happens if the first worker in~\cref{ex:repartitioning2} starts by processing $\bX_{4:6}$ after re-partitioning). Hence, when changing the number of subpartitions of the $i$-th worker from $p_i$ to $p_i'$, instead of using~\cref{eq:index_update}, we update $k_i$ using~\cref{al:alignment}, which relies on the function
\begin{equation} \notag
  p_\mathsf{trans}(n_i, p_i, p'_i, k_i) = \left\lceil p_\mathsf{start}(n_i, p_i, k_i) \frac{p_i'}{n_i} \right\rceil,
\end{equation}
that returns the index of the partition containing sample $p_\mathsf{start}(n_i, p_i, k_i)$ when the number of partitions is $p'_i$. We illustrate~\cref{al:alignment} with~\cref{ex:repartitioning3}.
\begin{example}[Continuation of~\cref{ex:repartitioning2}] \label{ex:repartitioning3}
  Say that, prior to re-partitioning, the first worker processed partition $\bX_{1:5}$, so that $k_1=1$, and that we are changing the number of subpartitions from $p_1=2$ to $p_1'=3$. In this case, the $n_1=10$ samples stored by the first worker are subpartitioned as follows, 
  \begin{equation} \notag
  \begin{aligned}
    p_1 = 2:\quad&& [1,\quad& 2,& 3\;,\quad & \;4, & 5],\quad& [6\;, & \;7,\quad& 8, & 9,\quad& 10]\\
    p_1' = 3:\quad&& [1,\quad& 2,& 3],\quad & [4, & 5,\quad& \;6], & [7,\quad& 8, & 9,\quad& 10]
  \end{aligned}
  \end{equation}  
  where the indices are the row indices of $\bm X$, and brackets in the first and second line indicate partition boundaries before and after re-partitioning, respectively.   
  Now,~\cref{al:alignment} finds a partition out of $p_1'=3$ partitions such that its first sample is equal to that of some partition out of $p_1=2$. It proceeds as follows. First, let $k_1 \leftarrow \mathrm{mod}\left(1, 2 \right) + 1 = 2$ (\cref{alg:line:1}), and $k_1' \leftarrow p_\mathsf{trans}(10, 2, 3, k_1) = 2$ (\cref{alg:line:2}). Since the $k_1$-th and $k_1'$-th partitions are not aligned (\cref{alg:line:3})\textemdash $p_\mathsf{start}(10, 3, k_1') = 4 \neq 6 = p_\mathsf{start}(10, 2, k_1)$\textemdash we let $k_1' \leftarrow k_1' - 1 = 1$ and $k_1 \leftarrow p_\mathsf{trans}(10, 3, 2, k_1') = 1$ (\cref{alg:line:4,alg:line:5}). Now the partitions are aligned (\cref{alg:line:3})\textemdash $p_\mathsf{start}(10, 2, k_1) = 1 = p_\mathsf{start}(10, 3, k_1')$\textemdash and the worker assigns $p_1 \leftarrow p_1'$ and $k_1 \leftarrow k_1'$ (\cref{alg:line:7,alg:line:8}).
\end{example}%
Note that~\cref{al:alignment} always terminates, since the first partition always starts at the first sample stored by the worker, i.e., $k_i = k_i' = 1$ results in the partitions being aligned regardless of the values of $p_i$ and $p_i'$. However, $k_i = k_i' = 1$ may not be the only solution. For example, if $n_i=10$, $p_i=2$, and $p_i'=4$, then $k_i=2$ and $k_i'=3$ also results in aligned partitions\textemdash $p_\mathsf{start}(10, 4, 3) = 6 = p_\mathsf{start}(10, 2, 2)$. Hence,~\cref{al:alignment} improves timeliness, since always setting $k_i = k_i' = 1$ after re-partitioning could result in the first few subpartitions being processed much more frequently than the others.

\section{Convergence results}
\label{sec:convergence}

\begin{figure*}[t]
  \centering
  \input{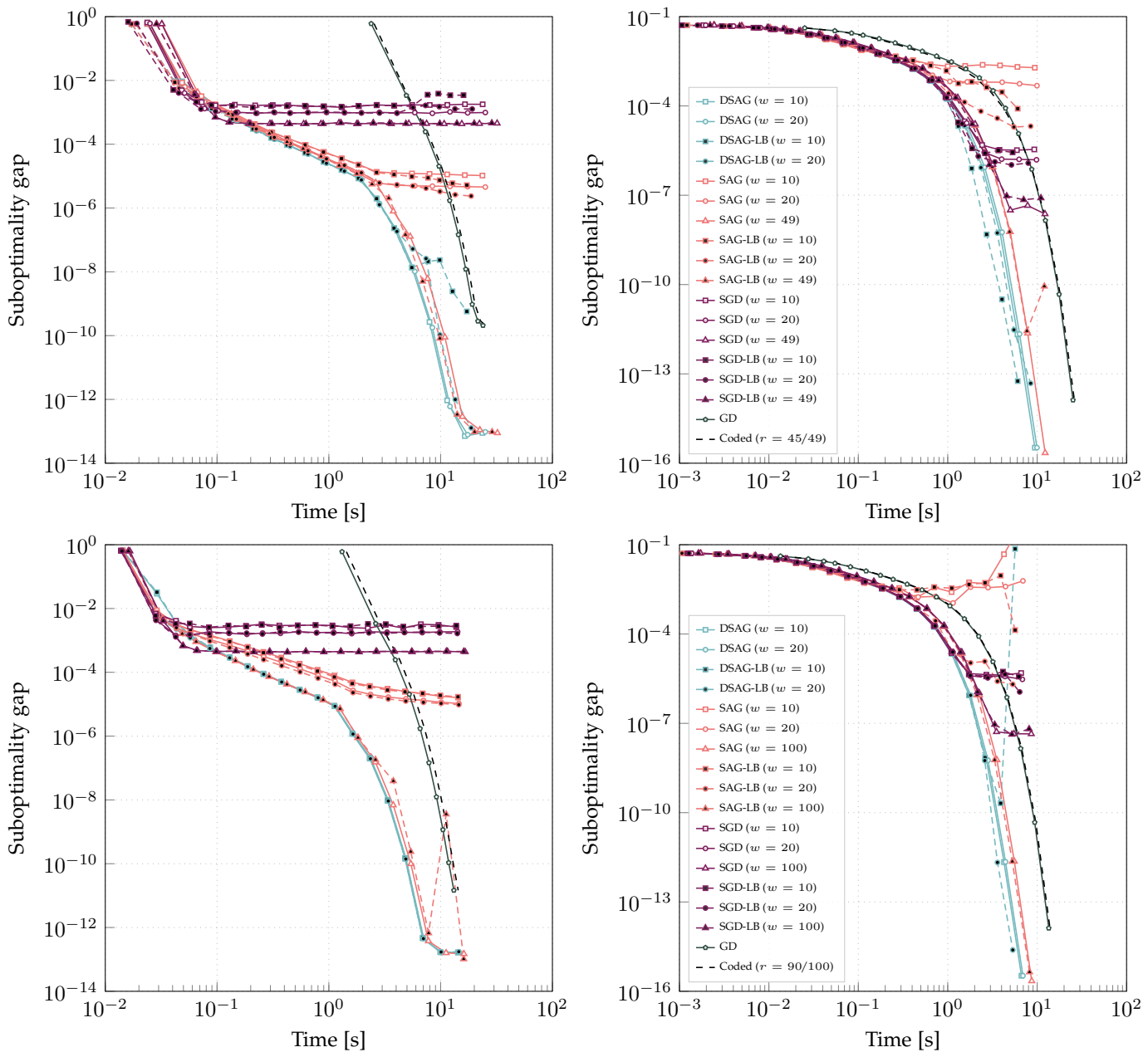}
  \caption{Convergence of PCA (left column) and logistic regression (right column) for $\Nn=49$ workers on eX3 (top row) and $\Nn=100$ workers on AWS (bottom row). The dataset is split evenly over the workers and is initially subdivided into \num{100} subpartitions for PCA and \num{10} subpartitions for logistic regression. Stochastic optimization methods with $w < \Nn$ effectively reduce the impact on latency of straggling workers, but only DSAG ensures convergence to the optimum. Load-balancing can improve latency further in some instances. The results shown are averages over \num{5} experiments.}
  \label{fig:convergence}
  \vspace{-5mm}
\end{figure*}

Here, we evaluate the performance of DSAG for PCA and logistic regression, and compare it to that of GD, SGD, SAG, and coded computing methods, on eX3 and AWS (see~\cref{sec:computing} for details). We also evaluate the impact of load-balancing on performance   for DSAG, SAG, and SGD. For PCA, the loss function is given by
\begin{equation} \label{eq:pca}
  R(\bV) = \frac{1}{2} \left\lVert \bV \right\rVert_\mathsf{F}^2
  \;\text{ and }\;
  f_i(\bV) = \frac{1}{2} \left\lVert \bx_i - \bx_i \bV \transpose{\bV} \right\rVert^2,
\end{equation}
where the columns of $\bV$ make up the computed principal components, $\lVert \cdot \rVert$ denotes the Euclidean norm, and $\|\cdot\|_\mathsf{F}$ denotes the Frobenius norm and $\bV$ is updated according to \eqref{eq:pcagd}. For PCA, $G(\cdot)$ in \eqref{eq:pcagd}  is the Gram-Schmidt operator, i.e., $G(\cdot)$ takes an input matrix and applies the Gram-Schmidt orthogonalization procedure to its columns such that the columns of the resulting matrix form an orthonormal basis with the same span as the columns of the input matrix.
For logistic regression, the loss is the \emph{L2-regularized} classification error, i.e.,
\begin{equation} \notag
  R(\bV) = \frac{\lambda}{2}\lVert \bm{V} \rVert^2
  \;\text{ and }\;
  f_i(\bV) = \frac{\log\left[ 1 + \exp\left( -b_i \transpose{\bx_i} \bV \right) \right]}{n},
\end{equation}
where $b_1, \dots, b_n$ are the classification labels, with $b_i \in \{-1, +1\}$, $\lambda$ is the regularization coefficient, and in this case $G(\cdot)$ is the identity operator. For PCA, we use a matrix derived from the $1000$ Genomes phase-$3$ dataset~\cite{Auton2015}, and for logistic regression we use the HIGGS dataset~\cite{Baldi2014} (see~\cref{sec:computing}). For PCA, we compute the top \num{3} principle components, and for logistic regression, as in~\cite{Schmidt2017}, we normalize all features to have zero mean and unit variance, add an intercept equal to \num{1}, and set the regularization coefficient to \num{1} divided by the number of samples, i.e., $\lambda = 1/\num{11 000 000}$. We use \num{100} and \num{10} subpartitions for PCA and logistic regression, respectively.

We measure performance as the latency to solve either PCA or logistic regression to within some precision of the optimum, and, for all scenarios, we plot the suboptimality gap, i.e., the difference between the explained variance (for PCA) or classification error (for logistic regression) of the computed solution and that of the optimum, as a function of time. The results shown are averages over \num{5} experiments conducted on the respective computing systems.
For GD and coded computing, we use a stepsize of $\eta=1.0$ for both PCA and logistic regression, whereas for DSAG, SAG, and SGD, we use a stepsize of $\eta=0.9$ for PCA and $\eta=0.25$ for logistic regression (we need to reduce the stepsize relative to GD for the stochastic methods to ensure convergence). We remark that GD applied to solving the optimization problem in~\eqref{eq:optimization} with the loss function in~\cref{eq:pca} with $\eta=1.0$ is equivalent to the \emph{power method} for PCA, i.e., the power method is a special case of GD.

\subsection{Coded computing}

Coded computing methods with \emph{code rate} $r$ (a quantity between \num{0} and \num{1}) make it possible to either recover the gradient exactly (e.g.,~\cite{Tandon2017}) or an approximation thereof (e.g.,~\cite{Karakus2017,Wang2019,Bitar2020,Yang2018}) from intermediate results computed by a subset of the workers, at the expense of increasing the computational load of each worker by a factor $1/r$ relative to GD. The gradient is recovered via a decoding operation (that typically reduces to solving a system of linear equations), the complexity of which usually increases superlinearly with the number of workers. Ideally, the gradient can be recovered exactly from the results computed by any set of $\left\lceil r \Nn \right\rceil$ workers\textemdash codes with this property are referred to as \emph{maximum distance separable} (MDS) codes\textemdash but increasing the number of results required can allow for reducing the decoding complexity~\cite{Severinson2019tcom}.

To compare against the wide range of coded computing methods, we use an idealized estimate derived from the GD results. In particular, we assume that the code is MDS, but that the decoding complexity is zero. More specifically, we set the latency per iteration equal to that of the $\left\lceil r \Nn \right\rceil$-th fastest worker after scaling the computational latency recorded for GD of all workers by $1/r$, and the rate of convergence equal to that of GD. Hence, both the latency and rate of convergence of the estimate are bounds on what is achievable with coded computing. Further, for PCA, this bound includes coded computing methods for matrix multiplication (e.g.,~\cite{Lee2017,Severinson2019tcom,Severinson2018,Yang2018}), since GD is equivalent to the power method in this instance.

\subsection{Artificial scenario}

While we are primarily interested in cloud computing systems, for the sake of reproducibility, we first present results recorded for $\Nn=49$ workers on eX3, which is much more homogenous than the cloud, where we introduce variability in a controlled manner. In particular, we artificially increase the computational latency of the $i$-th worker by a factor $(i/\Nn) \cdot \num{0.4}$ by introducing delays at the worker nodes.\footnote{This level of variability is comparable to what we have observed for instances of type \texttt{F2s\_v2} on Azure.} Further, we remove this artificial latency for workers \num{40} through \num{49} after one second has passed from the start of the learning process to simulate those workers coming out of a high-latency burst.

In~\cref{fig:convergence} (top row), we show convergence of PCA (left) and logistic regression (right) in this scenario. First, for both PCA and logistic regression, at least one of the stochastic methods (DSAG, SAG, and SGD) is more than twice as fast as GD for any suboptimality gap\textemdash performing many fast, but inexact iterations, is often preferable to performing fewer more accurate iterations. However, for SAG, when $w < \Nn$, and SGD, there is a point beyond which convergence effectively stops. For SGD, the high variance of its gradient estimate prevents it from converging\textemdash SGD is not a variance-reduced method\footnote{A popular variance reduction technique for SGD is to gradually decrease the stepsize, but doing so reduces the rate of convergence.}\textemdash although larger $w$ increases precision since it causes a larger fraction of the dataset to be factored in. For SAG, which is variance-reduced, convergence stops as a result of not factoring in samples stored by workers that are straggling over many subsequent iterations (see~\cref{sec:dsag_dist}). For $w=\Nn$, SAG converges to the optimum since all workers participate in each iteration, at the expense of increased latency, i.e., there is a trade-off between straggler-resiliency and convergence.

DSAG extends SAG by incorporating stale results, and, as a result, converges to the optimum even when $w < \Nn$, allowing it to achieve both low latency and high precision in the presence of stragglers. In this instance, DSAG with $w=10$ is the fastest  of all methods considered for both PCA and logistic regression, except for when solving PCA to within a precision of about $10^{-3}$, in which case SGD is faster. In particular, DSAG with $w=10$ achieves a rate of convergence comparable to that of SAG with $w=\Nn$, but reduces latency by an amount that is proportional to the amount of latency variability. For example, for PCA, DSAG with $w=10$ is between about \num{20}\% (for a suboptimality gap of $10^{-4}$) and \num{30}\% (for a suboptimality gap of $10^{-8}$ or lower) faster than SAG with $w=\Nn$, and, for logistic regression, DSAG with $w=10$ is about \num{30}\% faster than SAG when the suboptimality gap is $10^{-4}$ or lower. Finally, for both PCA and logistic regression, the straggler resiliency afforded by coding is canceled out by the higher computational load. Here, we consider a code rate $r=45/49$, which we find yields lower latency compared to the lower rates typically used in coded computing (e.g., in~\cite{Tandon2017,Karakus2017,Wang2019,Bitar2020,Yang2018}).

Next, we evaluate the proposed load-balancer, which we apply to DSAG, SAG, and SGD\textemdash we refer to the corresponding load-balanced methods as DSAG-LB, SAG-LB, and SGD-LB, respectively. For SAG-LB, to allow for dynamically re-sizing the data partitions, we use the DSAG update rule (see~\cref{sec:dsag}), except that stale results are discarded, instead of that in~\cite{Schmidt2017}. There are two important caveats. First, it takes about \num{7} and \num{0.5} seconds for the load-balancer to produce a first solution for PCA and logistic regression, respectively, before which it has no effect (it is slower for PCA due to the larger number of subpartitions). Second, load-balancing can reduce precision when the suboptimality gap is low due to cache invalidation (see~\cref{ex:repartitioning}).\footnote{This problem could be alleviated by disabling load-balancing when close to convergence.} This problem is especially severe when the number of subpartitions is large relative to the total number of iterations (as is the case for the PCA problem we consider) since a larger fraction of the overall optimization time is spent before the cache is re-populated. As a result, load-balancing does not result in a speedup for PCA. However, for DSAG with $w=10$ applied to logistic regression, load-balancing results in about \num{30}\% to \num{40}\% lower latency when the suboptimality gap is between $10^{-6}$ and $10^{-12}$. Interestingly, the primary mechanism by which load-balancing reduces latency is by increasing the average number of workers that respond within the \num{2}\% latency tolerance (see~\cref{sec:dsag_dist}), which allows it to reduce the workload for all workers without reducing the expected overall contribution (see~\cref{sec:optimizer}). Further, load-balancing improves the precision of SAG with $w<\Nn$ since the probability of each worker participating becomes more uniform.


%
%

\subsection{Performance on AWS}

\ctable[
cap = ,
caption = Approximate latency of stochastic methods.,
label = tab:latency,
pos = t
]{lll}{}{ \FL
& Comm. latency [s] & Comp. latency [s] \ML
\hspace{-1mm}eX3 PCA\hspace{-3mm} & \numrange{2e-5}{6e-5}\hspace{-2mm} & \numrange{2.2e-2}{3.1e-2}\hspace{-1mm} \NN
\hspace{-1mm}AWS PCA\hspace{-3mm} & \numrange{1.5e-4}{1e-3}\hspace{-2mm} & \numrange{1.3e-2}{1.6e-2}\hspace{-1mm} \NN
\hspace{-1mm}eX3 Log. r.\hspace{-3mm} & \numrange{0.2e-5}{3e-5}\hspace{-2mm} & \numrange{1.8e-3}{2.5e-3}\hspace{-1mm} \NN
\hspace{-1mm}AWS Log. r.\hspace{-3mm} & \numrange{1e-4}{6e-4}\hspace{-2mm} & \numrange{1.1e-3}{1.3e-3}\hspace{-1mm} \LL
\vspace{-5mm}
}

Here, we consider performance on a cluster composed of $N=\num{100}$ workers on AWS. To ensure that the results are representative, we use a fresh set of virtual machine instances for each set of experiments. While the results on AWS are similar to those on eX3, there are a few important differences. First, communication latency is about an order of magnitude higher on AWS compared to eX3, whereas computation latency is about \num{10}\% to \num{30}\% higher, depending on the scenario (when accounting for the fact that the per-worker computational load is about half that of eX3). We show the approximate latency range for the stochastic methods without load-balancing in~\cref{tab:latency}. As a result, the performance advantage of the stochastic methods compared to GD and coded computing is reduced somewhat, although they are still about twice as fast.

Second, latency is noisier on AWS, with workers experiencing unpredictable high-latency bursts, which may affect both communication and computation latency. Further, the noise makes up a larger fraction of the overall latency for lower average latency. As a result, the straggler problem is more severe for logistic regression than for PCA, for which each iteration is much slower (see~\cref{tab:latency}). In particular, for PCA, DSAG with $w=10$ is only up to about \num{10}\% faster than SAG with $w=\Nn$ (for a suboptimality gap below $10^{-6}$), whereas for logistic regression DSAG with $w=5$ is about \num{30}\% faster when the suboptimality gap is $10^{-4}$ or lower.

Finally, the level of static variation in latency between workers is smaller on AWS than on eX3 (which we modeled after Azure). Hence, the advantage of load-balancing is smaller\textemdash about \num{10}\%  to \num{15}\% for DSAG-LB with $w=20$ compared to DSAG with $w=10$ (which is fastest when not load-balancing), for logistic regression, and up to about \num{50}\% faster than SAG with $w=\Nn$.

%
%
%

\section{Conclusions}

Recently, there has been significant interest in coded computing, which is often motivated by the straggler problem in distributed machine learning and data analytics. However, we find that there are applications for which coded computing reduces performance compared to GD, even when not accounting for the decoding latency, which may be substantial. One issue is that coded computing methods are often designed under the assumption that latency is i.i.d. between workers, which is typically not the case. Further, there are fundamental differences between the distributed computing problem and the communication problem that erasure correcting codes were designed to address. In particular, we find that, for iterative methods, missing information can be substituted by stale information received over previous iterations, with only a marginal reduction to the rate of convergence. In this way, variance-reduced stochastic optimization methods can achieve straggler resiliency without increasing computational complexity, as is the case for coded computing.

In this work, we have proposed DSAG, which alleviates the straggler problem by only waiting for the fastest subset of workers, while integrating the results computed by stragglers in an asynchronous manner. DSAG is based on the SAG method and uses a variance reduction strategy to improve convergence. Further, we have proposed a load-balancing strategy that is able to counter some of the latency variability that exists in distributed computing systems, without moving data between workers. For both PCA and logistic regression, we have shown that DSAG can reduce latency significantly\textemdash by up to \num{50}\% for logistic regression on AWS, compared to SAG\textemdash through a combination of load-balancing and only waiting for the fastest subset of workers.

\balance
\bibliographystyle{IEEEtran}
\bibliography{./manuscript}

\end{document}